\documentclass[
 aip,
 jp,
amsmath,amssymb,
preprint,
noeprint]{revtex4-1}

\usepackage{physics}
\usepackage{graphicx}
\usepackage{dcolumn}
\usepackage{bm}
\usepackage{verbatim} 
\usepackage{mathtools}
\usepackage{natbib}
\usepackage{bbold}
\usepackage[version=4]{mhchem}
\usepackage{amsmath}
\usepackage{accents}
\usepackage{xcolor}
\usepackage{color,soul}

\begin{document}
\raggedbottom

\title{A Grid-free Approach for Simulating Sweep and Cyclic Voltammetry.}

\author{Alec J. Coffman}
\affiliation{Department of Chemistry, University of Pennsylvania, Philadelphia, Pennsylvania 19104, USA}
\author{Jianfeng Lu}
\affiliation{Departments of Mathematics, Physics, and Chemistry, Duke University, Durham, North Carolina 27708, USA}
\author{Joseph E. Subotnik}
\affiliation{Department of Chemistry, University of Pennsylvania, Philadelphia, Pennsylvania 19104, USA}

\date{\today}

\begin{abstract}
We present a new computational approach to simulate linear sweep and cyclic voltammetry experiments that does not require a discretized grid in space to quantify diffusion. By using a Green's function solution coupled to a standard implicit ordinary differential equation
solver, we are able to simulate current and 
redox species concentrations using only a small grid in \textit{time}. As a result,
where benchmarking is possible, we find that the current method is faster (and quantitatively identical) to established techniques. The present algorithm should help open the door to
studying adsorption effects in inner sphere electrochemistry. 

\end{abstract}

\maketitle

\section{Introduction}
\subsection{Models for Outer-Sphere Electrochemical electron transfer (ET) and Voltammetry}

A common experimental technique throughout the field of electrochemistry is sweep voltammetry, where the electrostatic potential of a metal electrode (relative to an electrochemical cell containing an electroactive redox species)  is changed with time. 
The resultant current is usually plotted as a function of this change in 
driving force\cite{Elgrishi2018}.
When the potential is ramped in a linear fashion, $E(t) = E_{0} + \nu t$, 
(where $\nu$ is the scan rate), the resulting plot is known as a linear sweep voltammetric (LSV) 
curve.
When the scan is carried out in both directions (oxidation and reduction), it is known as cyclic voltammetry (CV).
The current in such experiments arise from both (i) the change in the electron transfer (ET) kinetics at the electrode (due to changing the driving force for the ET reaction), and also (ii) diffusional effects which create a concentration gradient of the redox active species in solution.
This novel combination of effects driving the current vs.\ potential profile makes linear sweep voltammetry an ideal experiment for inferring unknown chemical and physical parameters of interest in the electrochemical system. In particular, nowadays, it is routine to use LSV to infer
the diffusion constant of the redox active species\cite{Bogdan2014} or the standard rate constant for an ET reaction, $k^{0}$ (the rate constant where the forward and backward rate for the redox reactions are equal)\cite{Randviir2018}.

In order to properly analyze an LSV experiment, one requires a mathematical model with a reasonable description of both the diffusional and ET effects.
Consider the most basic electrochemical reaction, the reduction of a redox species $A$ to 
its reduced form $B$,\\
\centerline{\ce{A + e^- <=>[\ce{k_{f}}][\ce{k_{b}}] B}}. \\
% WTF \noindent where $A$ and $B$ can be either in solution or adsorbed to an electrode surface, depending on the instantiation of the problem.
where species $A$ exists in bulk solution but, before a reaction, must be brought to the
metal surface.
For most simulations, diffusional effects are modeled through a standard parabolic, second order partial differential equation (PDE)\cite{Bard2001}, 
\begin{equation}\label{eq:ODE_ex}
\begin{split}
&\frac{\partial c_{A}}{\partial t} = D_{A} \frac{\partial^{2} c_{A}}{\partial x^{2}}  \\
&\frac{\partial c_{B}}{\partial t} = D_{B} \frac{\partial^{2} c_{B}}{\partial x^{2}}.
\end{split}
\end{equation}

\noindent Thereafter (and see Eq.\ \ref{eq:ODE} below), ET effects are incorporated through either a source term in the PDE or electrode boundary conditions.
This approach has wide spread applicability and  can be adadpted to take into account electrical double layer (EDL) effects (which are also called Frumkin corrections to the rate constant that account for uncompensated resistance due to field drop at the point/plane of ET)\cite{VanSoestbergen2012, Limaye2020}, surface adsorption of a species pre-ET\cite{Lin2015, Samin2016, Yang2019, Yang2020}, coupled homogenous chemical reactions\cite{Shea1987, Bott1999, Chen2020}, and many other effects as well.
Some of the relevant PDEs can be solved analytically if the ET kinetics are considered to be infinitely fast (i.e. when ET is taken to be instantaneous and dependent upon only the thermodynamics of the redox couple at a given external potential); this is known as the Nernstian limit (again, see below).
Realistically, however, most ET kinetics are not infinitely fast and the Nernstian limit is an idealization. 
In practice, in order to interpret his/her data, the experimentalist will usually need to model voltammetry experiments explicitly, i.e.\ one requires a computer to solve the necessary PDEs (e.g., in Eq.\ \ref{eq:ODE}) that incorporate (at minimum) both diffusion and a time-dependent ET rate constant.

Finally, for completeness, a few words are appropriate as far as how most
electrochemical simulations model electron transfer if one wishes to go beyond the Nernstian limit.
In general,  there are two common choices for the rate constant: 
Butler-Volmer (BV) and Marcus-Hush-Chidsey (MHC).

\begin{enumerate}
\item
The most common choice for the ET rate constant is the Butler-Volmer (BV) expression.
BV dynamics are characterized by two parameters: the charge transfer coefficient, $\alpha$, and the standard prefactor $k^{0}$.
The charge transfer coefficient's physical interpretation has historically been a matter of some debate\cite{Guidelli2014}, but mathematically is defined as the change in the log of the current with respect to change in the driving force  (in unitless form):  $\alpha = \abs{\frac{RT}{F} \frac{d \ln{I}}{dE}}$.
The total BV forward ($k_{f}$) and backward ($k_{b}$) rate constants take the form 
\begin{equation}
\begin{split}
&k_{f} = k^{0} e^{-\alpha\frac{F}{RT}(E-E^{0'})} \\
&k_{b} = k^{0} e^{(1-\alpha)\frac{F}{RT}(E-E^{0'})},
\end{split}
\end{equation}
\noindent where $E^{0'}$ is the formal potential (which is dependent on the experimental setup), $F$ is the Faraday constant, and $R$ is the universal gas constant.

\begin{figure}[!htb]
\includegraphics[width=11cm,height=11cm]{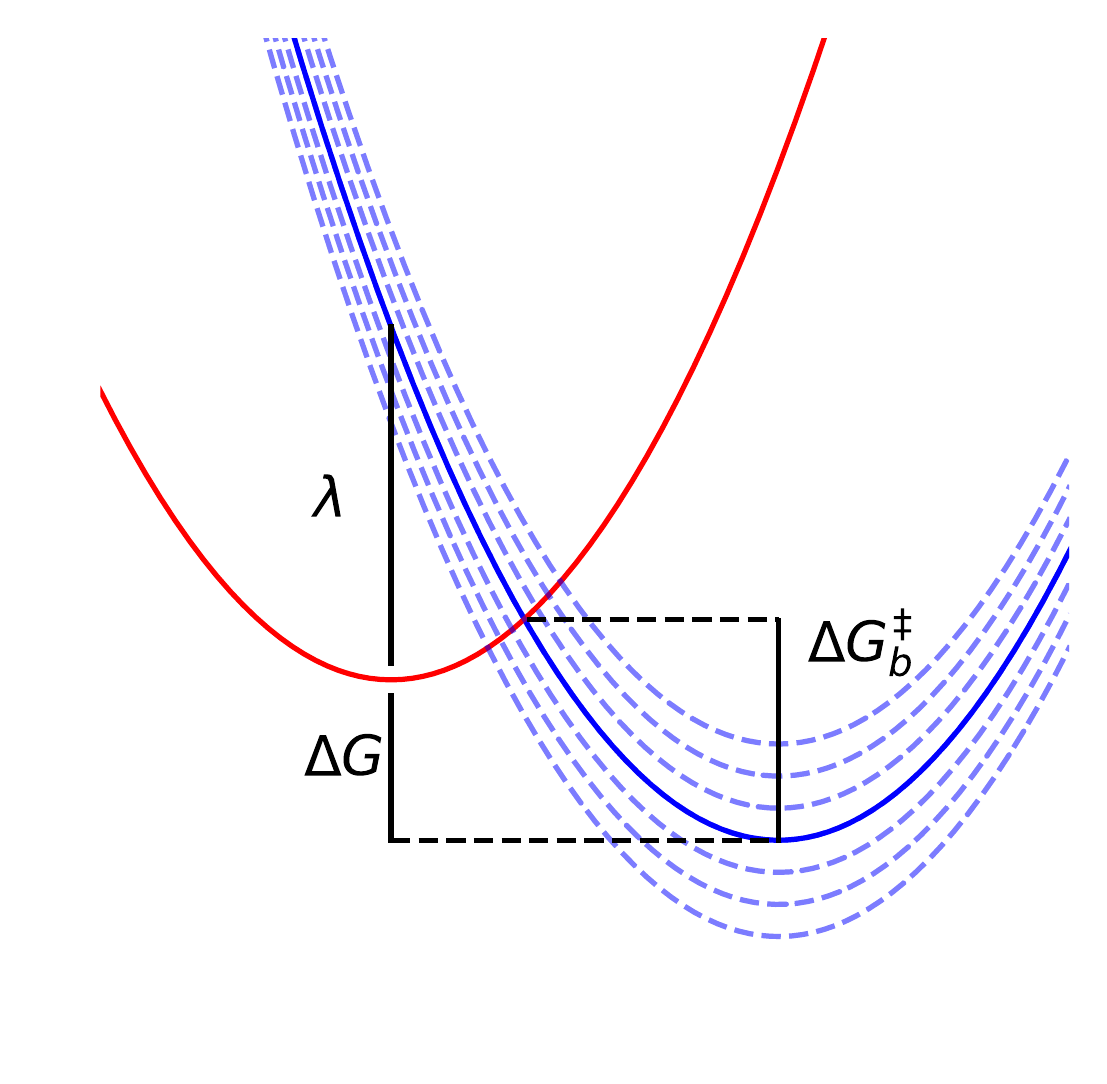}
\caption{\label{Figure:marcus_curve} The standard harmonic potential energy surfaces over a general solvent reorganization coordinate within the Marcus picture.
The red curve represents the charge located on a redox species molecule, while the blue curves represent the charge being located on one of the continuum of possible electronic states that are accessible in the metal electrode.
The free energy difference between the charge being located on the molecule versus the metallic state at the fermi level of the metal electrode is given by $\Delta G$, while the reorganization energy ($\lambda$) is the energy required to change the relaxed conformation of the redox active species from a local solvent environment of reactant state to the local solvent environment in the product state (all while keeping the charge fixed in the product state).
The barrier for transferring an electron from the product to reactant state is also shown, labeled by $\Delta G_{b}^{\ddag}$.}
\end{figure}

\item
The less common form for the ET rate constant is the MHC expression.
Like BV, MHC can also
be characterized by two parameters, the prefactor $k^{0}$, 
and the reorganization energy $\lambda$. 
While the prefactor $k^{0}$ has the same interpretation as in BV theory,
the reorganization energy $\lambda$
is defined as the free energy change associated with relaxing
the local solvent environment from one configuration (equilibrated to the reactant state)
to another configuration (equilibrated to the product state), all the while keeping
the molecule's electronic state fixed as the product.

\begin{itemize}
\item
For homogenous ET in solution, at high temperature, MHC theory is equivalent to transition state theory, such that:
\begin{eqnarray}
k = k^{0} e^{-\frac{\Delta G^{\ddagger}}{k_{B}T}}
\end{eqnarray}
where $\Delta G^{\ddagger}$ is the barrier height associated with the transition state, while $k_{B}$ and $T$ have their usual definitions as the Boltzmann constant and temperature, respectively .
While the exact form of $k^0$ can be complicated and depend on the nature of the reaction (inner or outer sphere)\cite{Schmickler2010}, such a distinction will not be relevant for our present purposes.  In practice, if one makes a parabolic free energy assumption,
the Marcus rate takes on the famous form:
\begin{eqnarray}
k = k^{0} e^{\frac{(\lambda + \Delta G)^{2}}{4k_{B}T}},
\end{eqnarray}

\noindent where $\Delta G$ is the free energy difference between the electron being located on the acceptor versus the donor.
\item
For a heterogeneous electrochemical ET, the Marcus rate constant must be modified to account for the fact that the product is no longer a chemical species in solution, but rather a metal electrode with a continuum of electronic states, which obeys an occupancy dictated by a Fermi distribution at a given external potential.
By integrating the rate constants over the distribution of electronic states in the electrode, one obtains the so-called MHC rate constant for electrochemical ET,

\begin{equation}
\begin{split}
k_{f} = k^{0} \int_{-\infty}^{\infty} f(\epsilon) e^{-\frac{\Delta G_{f}^{\ddagger} (\epsilon)}{RT}} d\epsilon \\
k_{b} = k^{0} \int_{-\infty}^{\infty} f(-\epsilon) e^{-\frac{\Delta G_{b}^{\ddagger} (\epsilon)}{RT}} d\epsilon,
\end{split}
\end{equation}

\noindent Here $f(\epsilon) = \frac{1}{1 + e^{\beta\epsilon}}$ is the fermi function and $\beta = \frac{1}{k_{B}T}$.

\end{itemize}
\end{enumerate}

\subsection{Adsorptive Effects in Linear Sweep Voltammetry-A First Step Towards Inner-Sphere ET}

The above description of ET in LSV experiments is usually sufficient for ET that occurs in solution between redox species. However, if ET occurs between redox species that are adsorbed to the surface of the electrode, many new factors are introduced that complicate the mathematics.
These factors include (i) possible changes in the driving force for ET due to the adsorbed species being located within the electrical double layer (EDL); (ii) possibly different reaction mechanisms
that can affect both the kinetics and thermodynamics, i.e. inner vs. outer sphere, adiabatic vs. nonadiabatic mechanisms; (iii) the introduction of an additional timescale for mass transfer, 
i.e. one must now account not only for a species in the bulk approaching the surface,
but also for the rate of adsorption and desorption of the oxidized and reduced species
(as well as a diffusion times moving along the surface).
This list of important focuses at a surface is by no means exhaustive, but 
these three effects are likely the most important.

Historically, the numerical simulation of adsorption effects in voltammetry experiments has been challenging
because adsorption introduces nonlinearity into the differential equation in Eq.\ \ref{eq:ODE_ex}; see Eq.\ \ref{eq:ODE} below.
In order to conceptually separate nonlinear adsorption from the actual ET event, 
 electrochemists have developed a few different model isotherms
that quantify what concentration of a species $A$ is expected to be adsorbed on 
a given surface  $(\Gamma_{A})$
given a bulk concentration in solution $(c_A)$.
The most common adsorption isotherm is the Langmuir isotherm\cite{Langmuir1918}, 
which assumes that, at high enough solution concentrations, there is a saturating coverage of the electrode (denoted $\Gamma_{s}$).
The Langmuir isotherm further assumes that there are no interactions between the species on the electrode surface and that there is no surface heterogeneity.
Mathematically, for a single species $A$, the Langmuir isotherm is given by:
\begin{equation}\label{eq:Langmuir}
\frac{\Gamma_{A}}{\Gamma_{s} - \Gamma_{A}} = \beta_{A} c_{A}
\Longleftrightarrow
\frac{\Gamma_{A}}{\Gamma_{s}} = \frac{\beta_A c_A}{1 + \beta_A c_A},
\end{equation}

\noindent where $\beta_{A}$ is the equilibrium constant for the adsorption/desorption of $A$, $\beta_{A} = \frac{k_{ads}^{A}}{k_{des}^{A}}$.
For the case of multiple species (for example, $A$ and $B$, where $B$ is the reduced form of $A$), the surface concentration of species $i$ ($i = A,B$) can be expressed as 
\begin{equation}\label{eq:Langmuir_2}
\Gamma_{i} = \frac{\Gamma_{s}\beta_{i}c_{i}}{1 + \beta_{i}c_{i} + \sum\limits_{j \neq i} \beta_{j}c_{j}}
\end{equation}

\noindent More complicated isotherms are sometimes used in the literature,  such as the Frumkin isotherm\cite{Frumkin1964} (which allows for interactions between adsorbed species and the surface which can be coverage/concentration dependent). Below, we will present data from only the Langmuir isotherm (though the method itself will be quite general).

\section{Algorithm}
\subsection{Discretized Method}
Before presenting our grid-free approach, we quickly review
the standard approach for electrochemical voltammetry in the absence of adsorption.
Assuming that all electron transfer occurs only at the boundary,
the differential equation for the time evolution of the concentration of each species is:
\begin{equation} \label{eq:ODE}
\begin{split}
&\frac{\partial c_{A}}{\partial t} = D_{A} \frac{\partial^{2} c_{A}}{\partial x^{2}}  - (k_{f} c_{A} - k_{b} c_{B})\delta (x)  \\
&\frac{\partial c_{B}}{\partial t} = D_{B} \frac{\partial^{2} c_{B}}{\partial x^{2}}  + (k_{f} c_{A} - k_{b} c_{B})\delta (x) ,
\end{split}
\end{equation}
\noindent Here, $c_{A}$ and $c_{B}$ are the normalized concentrations of species $A$ and $B$, respectively, and $D_{A}$ and $D_{B}$ are the diffusion constants for species $A$ and $B$.
The delta function source term is used to enforce that ET occurs only at the solution-electrode interface, defined as $x=0$.

The equations above are typically solved on a discretized one-dimensional grid. In other words,
for a discrete set of points $\{x_1, x_2, ..., x_N\}$ with spacing $\Delta x$ and a time step $\Delta t$,
one solves for $c(j\Delta x, k\Delta t)$ in a big loop.
In principle, it is usually trivial to solve 
the differential equation in Eq.\ \ref{eq:ODE}
using the Euler method with a small enough time step. Unfortunately, however, the ET source terms  
at the metal electrode  (i.e. those terms proportional to $\delta(x)$)
usually lead to a myriad of numerical problems for any realistic time 
discretization (that would certainly dissuade most undergraduates).
Moreover, we cannot emphasize enough that electrochemistry occurs over a vast array of time scales.
While electron transfer may occur on the time scale of picoseconds, an electrochemical sweep over 1 V
with a scan rate of 10mV/s occurs over 100 seconds; one must propagate Eq. \ref{eq:ODE} over an immensely long time range.

The most common solution to such numerical difficulties\cite{Bard2001} 
is to transform the source terms in the above equation
into boundary conditions, utilizing a no-flux boundary condition at the electrode surface;
for details see Ref. \citenum{Coffman2019}. Thereafter, one must still
invoke an implicit ODE solver, as the equations are stiff\cite{Hindmarsh2005}.
Alternatively, because the operator on the right hand side of Eq.\ \ref{eq:ODE} is linear,
the simplest approach is to just diagonalize the corresponding (non-Hermitian) operator for diffusion and ET.
While diagonalization of a non-Hermitian operator is usually unstable,
it is well known that the non-Hermitian diffusion operator (in one dimension) can be transformed
into a Hermitian operator\cite{Risken1989}. As a result, if one is prepared to work with a grid
of points in one-spatial dimension, and one is not concerned with adsoprtion, there are a few numerically stable techniques available as far calculating
LSV curves.  Our recent work has shown that one can use such a brute
force approach to capture more complicated electrochemical scenarios, including
proton coupled electon transfer
\cite{Coffman2019, Coffman2020}.

\subsection{Grid-Free Method}
Unfortunately, the methods above face great difficulties when nonlinear adsorption and desorption are considered.  On the one hand,
 %These adsorption/desorption rates obtain their mathematical form from their respective isotherms; for the Langmuir isotherm, these are the expressions shown in Eqs.\ \ref{eq:Langmuir} and \ref{eq:Langmuir_2}.
%The non-linear aspects of these terms 
the presence of non-linear Langmuir isotherms lead to an even stronger stiffening
of the coupled set of PDEs, such that 
%which often prohibits the use of simple differential equation solving software in modeling the behavior of these systems.
using an implicit differntial equation solver to solve for the dynamics on a grid often fails.
On the other hand, diagonalization is obviously impossible with nonlinear equations of motion.
As a result, modeling adsorption is daunting for most parameter sets. 
The most successful algorithm to date comes from Compton {\em et al} who have used a spatial grid in a recent article to study phase transitions of adsorbed electroactive species\cite{Yang2020}. 

The goal of the present letter is to present a different solution to this problem, one that completely removes the necessity for a grid in space.  In so doing, we will vastly reduce the number of variables that must be propagated in time and, rather than spend time solving the heat equation in solution (and calculating the free solution concentration $c_A$), we will focus our attention exclusively at the surface where the electron transfer occurs (and calculate the surface concentration $\Gamma_A$).

\subsubsection{The relevant PDEs}
Before outlining our new approach, let us define the differential equations  for coupling diffusion, ET, and adsorption that we wish to solve. 
The PDEs for the solution concentrations ($c_A,c_B$) take the standard form (following Eq.\ \ref{eq:ODE} above), with the addition of adsorption/desorption rates, $\nu_{ads}$/$\nu_{des}$ at the electrode surface ($x=0$):
\begin{subequations} \label{eq:1D_sol}
\begin{equation}
\frac{\partial c_{A}}{\partial t} = D_{A} \frac{\partial^{2} c_{A}}{\partial x^{2}} - [(k^{sol}_{f} (t) c_{A} - k^{sol}_{b} (t) c_{B}) + \nu_{ads, A} - \nu_{des, A}]\delta (x) 
\end{equation}
\begin{equation}
\frac{\partial c_{B}}{\partial t} = D_{B} \frac{\partial^{2} c_{B}}{\partial x^{2}} + [(k^{sol}_{f} (t) c_{A} - k^{sol}_{b} (t) c_{B}) - \nu_{ads, B} + \nu_{des, B}]\delta (x). 
\end{equation}
\end{subequations}

\noindent The ODEs for the surface concentrations, $\Gamma_{A}$ and $\Gamma_{B}$ involve
both electron transfer as well as surface adsorption/desoprtion:
\begin{subequations} \label{eq:1D_surface}
\begin{equation}
\frac{\partial \Gamma_{A}}{\partial t} = \nu_{ads, A}(t) - \nu_{des, A}(t) - (k^{ads}_{f} (t) \Gamma_{A}(t) - k^{ads}_{b} (t) \Gamma_{B}(t))
\end{equation}
\begin{equation}
\frac{\partial \Gamma_{B}}{\partial t} =  \nu_{ads, B}(t) - \nu_{des, B}(t) + (k^{ads}_{f} (t) \Gamma_{A}(t) - k^{ads}_{b} (t) \Gamma_{B}(t)).
\end{equation}
\end{subequations}

\noindent The choice of isotherm will dictate the exact form of $\nu_{ads}^{i}/\nu_{des}^{i}$; e.g. for the case of a Langmuir isotherm, we choose 
\begin{subequations} \label{eq:ads_des}
\begin{eqnarray}
&\nu_{ads, A}(t)& = k_{ads}^{A} c_{A}(t,x=0) [\Gamma_{s} - (\Gamma_{A}(t) + \Gamma_{B}(t))] \\
&\nu_{des, A}(t)& = k_{des}^{A} \Gamma_{A}(t) \\
&\nu_{ads, B}(t)& = k_{ads}^{B} c_{B}(t,x=0) [\Gamma_{s} - (\Gamma_{A}(t) + \Gamma_{B}(t))] \\
&\nu_{des, B}(t)& = k_{des}^{B} \Gamma_{B}(t).
\end{eqnarray}
\end{subequations}

\noindent $k_{f}^{ads}/k_{f}^{sol} (k_{b}^{ads}/k_{b}^{sol})$ can follow different ET rate laws and are parameterized in different ways (different $\alpha/\lambda, k^{0}, E^{0}$, etc.) to reflect the correct underlying physics.

Lastly, when calculating the current, note that in principle there will be  separate contributions from both solution and adsorbed ET,
\begin{equation} \label{eq:current}
I(t) = nFA[(k_{f}^{sol}(t)c_{A}(t,x=0) - k_{b}^{sol}(t)c_{B}(t,x=0)) + (k_{f}^{ads}(t)\Gamma_{A} - k_{b}^{ads}(t)\Gamma_{B})],
\end{equation} 
\noindent where $n$ is the number of electrons, $A$ the surface area of the electrode, and $F$ is the Faraday constant.
Formally, the current depends only on four variables at a single time: $c_{A}(x=0, t), c_{B}(x=0, t), \Gamma_{A}(t),$ and $\Gamma_{B}(t)$.

\subsubsection{A Green's Function Approach}
Given that $\nu_{ads,i}/\nu_{des,i}$ and $I(t)$ depend only on $c_{A}(x=0), c_{B}(x=0), \Gamma_{A},$ and $\Gamma_{B}$ (and specifically not on any of the remaining $c_{A}(x \neq 0), c_{B}(x \neq 0)$), our goal is to find 
a maximally efficient algorithm that tracks only these four values
(without tracking the remaining solution concentrations away from the electrode surface).
Here, for simplicity, we will assume that species $A$ and $B$ have equal diffusivities ($D_{A} = D_{B}$); all of the following results can be simply modified to accommodate unequal diffusion constants, if needed.

To avoid using a grid in space, we invoke Duhamel's principle\cite{John1978} to recast our coupled differential equations modeling $c_{A}$ and $c_{B}$ as integral equations. 
For an inhomogeneous diffusion equation of the form Eq.\ \ref{eq:1D_sol}, with initial and boundary conditions $c_{B}(t=0, x) =\ c_{B}(t, x = \pm \infty) = 0$, $c_{A}(t=0, x) =\ c_{A}(t, x = \pm \infty) = 1$, one can formally invert
these equations of motion to solve for the solution concentration
in integral form (we will show $c_{B}$, but the derivation is identical for $c_{A}$):
\begin{equation}\label{eq:duhamel}
\begin{split}
c_{B}(t, x) =& \int_{0}^{t}\int_{0}^{x} G(x-y, t-s) [(k^{sol}_{f} (s) c_{A}(s, y) - k^{sol}_{b} (s) c_{B}(s,y)) - \nu_{ads, B} + \nu_{des, B}]\delta(y) ds\ dy \\
=& \int_{0}^{t}\int_{0}^{x} G(x-y, t-s) [b(s, y) + a(s)c_{B}(s, y)]\delta(y) ds\ dy \\
=& \int_{0}^{t} G(x, t-s) [b(s, y) + a(s)c_{B}(s, 0)] ds
\end{split}
\end{equation}

\noindent where 

\begin{eqnarray}
G(x-y, t-s) = \frac{1}{\sqrt{D_{B} \pi (t - s)}}e^{-\frac{|x - y|^{2}}{4(t-s)}}
\end{eqnarray}
is one half of the heat equation Greens function in one dimension\cite{John1978}.  
Note that a factor of 2 appears here because 
the original diffusion equation spans only the half-space $x \in [0, \infty)$.

In Eq.\ \ref{eq:duhamel}, the expression for $b(s, y)$ and $a(s)$ are obtained directly from reorganization of Eq.\ \ref{eq:1D_sol}; for a Langmuir isotherm, these expressions are 
\begin{equation}
\begin{split}
&b(s, y) = k_{f}^{sol}(s)c_{A}(s, y) + k_{des}^{B}\Gamma_{B}(s) \\
&a(s) = -(k_{b}^{sol}(s) + k_{ads}^{B}[\Gamma_{s} - (\Gamma_{A}(s) + \Gamma_{B}(s))]).
\end{split}
\end{equation}

\noindent Since we are only concerned with the solution concentration at $x=0$, the equation we wish to numerically solve is given by
\begin{equation} \label{eq:final_duhamel}
c_{B}(t, 0) = \int_{0}^{t} \frac{1}{\sqrt{D_{B} \pi (t - s)}} [b(s, 0) + a(s)c_{B}(s, 0)] ds.
\end{equation}

We face two challenges to employing Eq.\ \ref{eq:final_duhamel} to measure electrochemical currents: 
\begin{enumerate}
\item $b(s, 0)$ and $a(s)$ depend on $\Gamma_{A}$, $\Gamma_{B}$, and $c_{A}$ explicitly, while $\Gamma_{A}$ and $\Gamma_{B}$ depend explicitly on $c_{A}$ and $c_{B}$.
\item There is a singularity in Eq.\ \ref{eq:final_duhamel} at $s=t$, which must be carefully accounted for due to the importance of accurate integral evaluation at this point.
\end{enumerate}

Let us begin with the second challenge.  
To treat the singularity, we cast the integrand in Eq. \ref{eq:final_duhamel} into a general form of $\frac{f(s)}{\sqrt{t-s}}$, where $f(s) = b(s, 0) + a(s)c_{B}(s,0)$. 
We then break this term into two pieces; $\frac{f(s) - f(t)}{\sqrt{t-s}}$ and $\frac{f(t)}{\sqrt{t-s}}$.
The first piece goes to zero at $s=t$, and the second piece can be integrated analytically, removing the singularity.
By using nonweighted numerical quadrature (according to the trapezoid rule), one can obtain a numerical prescription for calculating $c_{B}(t)$ as a function of $b(s, 0)$, $a(s)$, and the prior history of $c_{B}$ (expressed on a grid $[s_{1}, s_{2},...,s_{N}]$, with spacing $\Delta s$):
\begin{equation} \label{eq:c_B_exact}
c_{B}(t, 0) = \frac{\Delta s (\sum\limits_{i=1}^{N-1} \frac{b(s_{i}, 0) + a(s_{i})c_{B}(s_{i}, 0) - b(t, 0)}{\sqrt{t-s_{i}}} + \frac{b(0, 0) - b(t, 0)}{2\sqrt{t}}) + 2b(t, 0)\sqrt{t}}{\sqrt{D_{B}\pi} - a(t)[2\sqrt{t}-\Delta s(\frac{1}{2\sqrt{t}} + \sum\limits_{i=1}^{N-1}\frac{1}{t - s_{i}})]}.
\end{equation} 

\noindent For a derivation, see Appendix I.
Note that, in the absence of adsorption, Eq.\ \ref{eq:c_B_exact} is entirely sufficient for calculating the current as a function of time/voltage for a sweep voltammetry simulation -- without any need for solving an ODE on a discretized grid in space.
For instance, suppose that $A$ and $B$ have equal diffusion constants for $A$ and $B$\footnote{
When $D_{A} \neq D_{B}$, then one must use the more strict condition that there is no net flux $D_{A} \frac{dc_{A}}{dx} = - D_{B} \frac{dc_{B}}{dx}$, which implies $D_{A} c_{A} = D_{B} c_{B}$ is a constant in space.}.
In such a case, $c_{A}(t, 0) + c_{B}(t, 0) = c_{A}(0, x = \infty) + c_{B}(0, x=\infty) = c_{bulk}$, so that $c_{A}(t, x=0)$ can be determined by $c_{A}(t, 0) = c_{bulk} - c_{B}(t, 0)$.
Therefore, since $k_{f}^{sol}$ and $k_{b}^{sol}$ are analytic expressions (already needed for evaluation of Eq.\ \ref{eq:c_B_exact}), one can easily calculate the current at time $t$ using Eq.\ \ref{eq:current} and \ref{eq:c_B_exact},
and thus challenge \#1 above disappears.
Implementing equation \ref{eq:c_B_exact} is as fast as (or faster than) any implicit ODE based scheme for evaluating equations of the type in Eq.\ \ref{eq:ODE}, and requires no black-box type assumptions that are often inherent when implicitly solving stiff ODEs with common numerical software packages\cite{Hindmarsh2005}.

Next, let us consider the first challenge, i.e.\ how to self consistently calculate concentrations when there is adsorption of redox active species.
In order to simultaneously solve Eq.\ \ref{eq:1D_surface} (a standard ODE) and Eq.\ \ref{eq:c_B_exact} (an integro-differential equation) in coupled fashion, we will apply the procedure below:
\begin{enumerate}
\item At $t=dt$, calculate $c_B(0, dt)$ using Eq.\ \ref{eq:c_B_exact} and the values of $\Gamma_{A}$, $\Gamma_{B}$ and $c_{A}$ at their prior time step values.
\item Calculate $c_{A}(0, dt)$ using the analogous Eq.\ \ref{eq:c_B_exact} for $c_{A}$, with the value of $c_{B}(0, dt)$ calculated in step 1 and the values of $\Gamma_{A}$ and $\Gamma_{B}$  at their prior time step values.
\item Return to step 1-2 until the tolerated convergence is achieved for $c_{A}(0, dt)$ and $c_{B}(0, dt)$.
\item Calculate $\Gamma_{A}(dt)$ and $\Gamma_{B}(dt)$ using Eq.\ \ref{eq:1D_surface} and an implicit ODE solver (in this case, backward differentiation formula of order 5).
\item Return to step 1 of this procedure until simultaneous convergence to a predetermined tolerance is achieved for all four values $c_{A}(0, dt)$, $c_{B}(0, dt)$, $\Gamma_{A}(dt)$, and $\Gamma_{B}(dt)$ within one single convergence step.
\item Move on to the next time step, repeating steps 1-5 at each time step until the end of the simulation is reached.
\end{enumerate}

\noindent Note that, in step 1) above, for the first guess of $\Gamma_{A}$ and $\Gamma_{B}$ at a given time step, extrapolation methods can be used to obtain a better initial guess than simply using the prior time step; this usually yields much faster convergence.
In our experience, linear extrapolation using two prior times ($\Gamma_{A}(t) = 2\Gamma_{A}(t - \delta t) - \Gamma_{A}(t - 2\delta t)$) yields a substantial improvement in convergence time, while higher order extrapolations do little for accelerating convergence.

\begin{figure}[!htb]
\includegraphics[width=11cm,height=13cm]{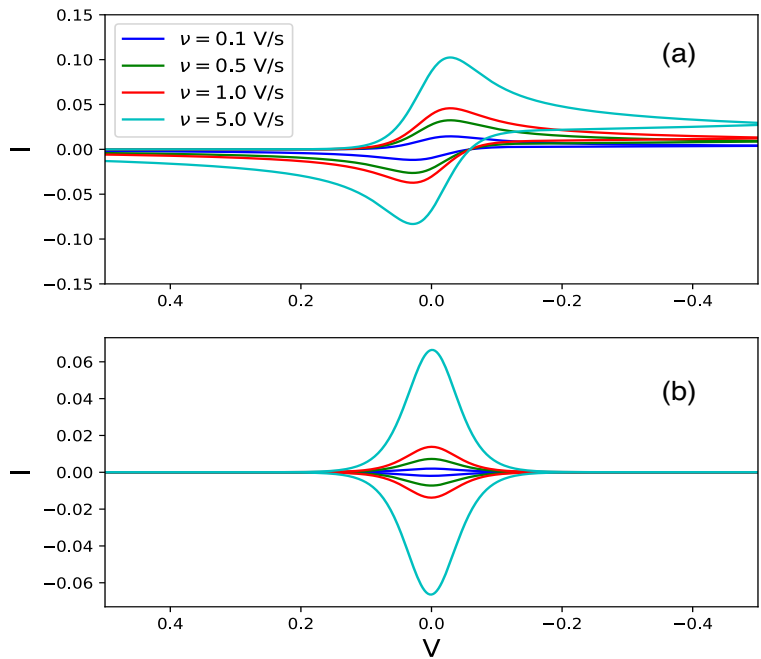}
\caption{\label{Figure:example_CV} Example CVs using the above method for at various scan rates, for the case with (a) no adsorption and (b) with adsorption.
The units for current here are $\frac{I}{nFA}$, while $V$ is in volts.
Mathematically, ``no adsorption'' means simulations using only Eq.\ \ref{eq:1D_sol} and not Eq.\ \ref{eq:1D_surface}, i.e.\ $\nu_{ads}$ and $\nu_{des}$ are set to zero.
``With adsorption'' implies using both Eqs.\ \ref{eq:1D_sol} and \ref{eq:1D_surface}, with $k_{f}^{sol}$ and $k_{b}^{sol}$ set to zero.
The parameters for these simulations are $D = 5.5 \cdot 10^{-4}$ cm\textsuperscript{2}/sec, $T = 300$ K, $N_{T} = 500$, $k_{0} = 10^{5}$ cm/sec, $\alpha = 0.5$ (BV kinetics are used for ET).
Wall times for each simulation can be found in Table 2.}
\end{figure}

In Fig.\ \ref{Figure:example_CV} we show a set of illustrative examples of CVs generated with this methodology, both in the case (a) without and (b) with adsorptive effects.
From the voltamogramms in Fig.\ \ref{Figure:example_CV}(a) without adsorption, at low scan rate we see the expected peak separation of $57 mV$ 
for a reversible ET reaction in solution, and the peak separation increases slightly as the scan rate increases and approaches the irreversible ET regime.
Conversely, in Fig.\ \ref{Figure:example_CV}(b), we see no peak separation in the case where ET occurs between surface bound species, as expected\cite{Bard2001}.
These results demonstrate that the above algorithm can capture the behavior of both coupled mass and electron transfer in solution and at surface bound electrochemical interfaces under a sweeping potential.

Additionally, to demonstrate the computational benefit of the present algorithm, below we show two tables illustrating the computational improvement of this method.
Table I shows computation times for the results shown in Fig.\ \ref{Figure:example_CV}, where we see that this present algorithm is relatively invariant to scan rate -- even with adsorption.
This behavior stands in contrast to the typical grid based ODE methods, which can require longer times at smaller scan rates, as more time is needed to propagate dynamics using an implicit ODE solver in order to cover a given potential range.
Lastly, Table II below shows the relative improvement in computation time for the present algorithm as compared to the more traditional approach where one solves the spatial grid-based coupled ODEs
through discretization of the relevant propagator over a set of grid points and diagonalization of the diffusion operator\cite{Coffman2019}.
Similar to the above argument about scan rate for traditional grid based methods, more spatial points implies larger computational cost, whereas for the method outlined in present paper, no such discretization is required as there is no added computational cost.
All simulations were performed in Python, using the NumPy and SciPy libraries, on a 2.8 GHz Dual-Core Intel Core i5 processor with 16 GB of RAM.

\begin{table}[ht]
\caption{Computational wall time by method (Python)}
\centering
\begin{tabular}{c c c}
\hline\hline
Scan rate (V/sec) & No adsorption & Adsorption \\ [0.5ex]
\hline
0.05 & 3.70 s & 22.06 s\\
0.1 & 4.89s & 21.13 s\\
1 & 3.40 s & 20.54 s\\
5 & 3.67 s & 19.87 s \\ [1ex] 
\hline 
\end{tabular}
\label{table:walltie}
\end{table}

\begin{table}[ht]
\caption{Computation time by method (Python)}
\centering
\begin{tabular}{c c}
\hline\hline
Algorithm & wall time \\ [0.5ex]
\hline
Grid-free &  2.34 s\\ 
Discretized, $N_{x} = 50$ & 9.94 s\\
Discretized, $N_{x} = 100$ & 30.91 s\\
Discretized, $N_{x} = 200$ & 83.02 s\\ [1ex] 
\hline 
\end{tabular}
\label{table:walltie}
\end{table}

\section{Conclusion}
In this article, as an alternative to traditional grid based methods, we have shown a novel integral based approach for simulating electrochemical ET to model ET along with adsorption of redox species.
Our approach is in quantitative agreement with grid based PDE solutions for a single ET experiment; more generally, however, the present algorithm should allow us to simulate more complicated sweep experiments that involve adsorption kinetics and beyond.
Code implementing this algorithm can be found at \url{https://github.com/alecja/electrochemistry_sim}.
Looking forward, we aim to apply this methodology to study the various parameter regimes of electrochemical voltammetry experiments where adsorption plays a key role.
After all, the addition of another timescale/parameter set corresponding to adsorption/desorption rates, plus the potential asymmetry in redox behavior of charge/uncharged species on a surface, 
implies that a realistic description of electrochemical dynamics and kinetics can and must go beyond the standard Nernstian and irreversible regimes that we have all grown up with.

\section{Acknowledgements}
This work was supported by the U.S. Air Force Office of Scientific Research (USAFOSR) under Grant Nos.\ FA9550-18-1-0497 and FA9550-18-1-0420, and the National Science Foundation under Grant No.\ DMS-2012286.

\section{Data Availability}
The data that support the findings of this study are available within the article.

\section{Appendix I: Derivation of Equation \ref{eq:c_B_exact}}
Below we will carry out a general derivation of Eq.\ \ref{eq:c_B_exact} from Eq.\ \ref{eq:final_duhamel}, for either species $A$ or $B$, with arbitrary initial conditions.
We first break the integrand into two pieces, $\frac{f(s) - f(t)}{\sqrt{t-s}}$ and $\frac{f(t)}{\sqrt{t-s}}$, where $f(s) = b(s, 0) + a(s)c_{j}(s,0)$.
The first integral is evaluated using the trapezoidal rule, while the second integral is evaluated analytically:
\begin{equation}
\begin{split}
c_{j}(t, 0) = \ &c_{j}(0,0) + (D_{j} \pi)^{-0.5} [\int_{0}^{t} \frac{[b(s, 0) + a(s)c_{j}(s, 0)]-[b(t, 0) + a(t)c_{j}(t, 0)]}{\sqrt{t - s}}ds + \\
&\int_{0}^{t} \frac{[b(t, 0) + a(t)c_{j}(t, 0)]}{\sqrt{t-s}} ds] \\
c_{j}(t, 0) = \ &c_{j}(0,0) + (D_{j} \pi)^{-0.5} [\Delta s (\sum\limits_{i=i}^{N-1} \frac{b(s_{i}, 0) + a(s_{i})c_{j}(s_{i}, 0) - b(t, 0) - a(t)c_{j}(t, 0)}{\sqrt{t-s_{i}}} + \\
&\frac{b(0, 0) - b(t, 0) + a(0)c_{j}(0, 0) - a(t)c_{j}(t, 0)}{2\sqrt{t}}) + 2 \sqrt(t) (b(t,0) + a(t,0)c_{j}(t,0))].
\end{split}
\end{equation}

\noindent Rearrangement of all $c_{j}(t, 0)$ terms to the left hand side yields 
\begin{equation}
\begin{split}
&c_{j}(t, 0) [\sqrt{D_{j} \pi} - 2 a(t)\sqrt{t} + \Delta s (\sum\limits_{i=1}^{N-1} \frac{a(t)}{\sqrt{t-s_{i}}} + \frac{a(t)}{2\sqrt{t}})] - c_{j}(0, 0) \sqrt{D_{j} \pi} = \\
&\Delta s(\sum\limits_{i=1}^{N-1} \frac{b(s_{i}, 0) + a(s_{i})c_{j}(s_{i}, 0) - b(t, 0)}{\sqrt{t-s_{i}}} + \frac{0.5(b(0, 0) - b(t, 0) + a(0)c_{j}(0,0))}{\sqrt{t}}) + 2b(t, 0)\sqrt{t}.
\end{split}
\end{equation}

\noindent Finally, isolation of $c_{j}(t,0)$ yields the final result,
\begin{equation}
\begin{split}
c_{j}(t, 0) = \frac{\Delta s(\sum\limits_{i=1}^{N-1} \frac{b(s_{i}, 0) + a(s_{i})c_{j}(s_{i}, 0) - b(t, 0)}{\sqrt{t-s_{i}}} + \frac{b(0, 0) - b(t, 0) + a(0)c_{j}(0,0)}{2\sqrt{t}}) + 2b(t, 0)\sqrt{t} + c_{j}(0, 0) \sqrt{D_{j} \pi}}{\sqrt{D_{j} \pi} - a(t)[2\sqrt{t} - \Delta s (\sum\limits_{i=1}^{N-1} \frac{1}{\sqrt{t-s_{i}}} + \frac{1}{2\sqrt{t}})]}
\end{split}
\end{equation}

\noindent Note that Eq.\ \ref{eq:c_B_exact} is recovered exactly when the index $j$ is replaced with $B$ and the initial condition $c_{B}(0,0) = 0$ is substituted.

\bibliographystyle{apsrev4-1}
\bibliography{alec_1}
\end{document}